\documentclass[sigchi]{acmart}

\usepackage{hyperref}
\usepackage{natbib}
\usepackage{booktabs} 
\usepackage[utf8x]{inputenc} 
\usepackage{balance}
\usepackage{titlesec}

\def\BibTeX{{\rm B\kern-.05em{\sc i\kern-.025em b}\kern-.08emT\kern-.1667em\lower.7ex\hbox{E}\kern-.125emX}}
\copyrightyear{2019}
\acmYear{2019}
\setcopyright{acmlicensed}
\acmConference[MobileHCI 2019]{MobileHCI 2019: 21st International Conference on Human-Computer Interaction with Mobile Devices and Services}{October 01--04, 2019}{Taipei, Taiwan}
\acmBooktitle{MobileHCI 2019: 21st International Conference on Human-Computer Interaction with Mobile Devices and Services, October 01--04, 2019, Taiwan, Taipei}
\acmPrice{15.00}
\acmDOI{10.1145/3338286.3340116}
\acmISBN{978-1-4503-6825-4/19/10}

\begin{document}

%
\title{Mapping Perceptions of Humanness in Speech-Based Intelligent Personal Assistant Interaction}
\fancyhead{}

%
\author{Philip R. Doyle}
\affiliation{University College Dublin}
\email{philip.doyle1@ucdconnect.ie}

\author{Justin Edwards}
\affiliation{University College Dublin}
\email{justin.edwards@ucdconnect.ie}

\author{Odile Dumbleton}
\affiliation{University College Dublin}
\email{odile.dumbleton@ucdconnect.ie}

\author{Leigh Clark}
\affiliation{University College Dublin}
\email{leigh.clark@ucd.ie}

\author{Benjamin R. Cowan}
\affiliation{University College Dublin}
\email{benjamin.cowan@ucd.ie}

%
\renewcommand{\shortauthors}{Philip R. Doyle et al.}

%
\begin{abstract}
Humanness is core to speech interface design. Yet little is known about how users conceptualise perceptions of humanness and how people define their interaction with speech interfaces through this. To map these perceptions n=21 participants held dialogues with a human and two speech interface based intelligent personal assistants, and then reflected and compared their experiences using the repertory grid technique. Analysis of the constructs show that perceptions of  humanness are multidimensional, focusing on eight key themes: \textit{partner knowledge set, interpersonal connection, linguistic content, partner performance and capabilities,  conversational interaction, partner identity and role, vocal qualities} and \textit{behavioral affordances}. Through these themes, it is clear that users define the capabilities of speech interfaces differently to humans, seeing them as more formal, fact based, impersonal and less authentic. Based on the findings, we discuss how the themes help to scaffold, categorise and target research and design efforts, considering the appropriateness of emulating humanness.  
\end{abstract}

%
%

%

%

%
\maketitle

\section{Introduction}
Speech interfaces like Amazon Alexa, Google Assistant and Apple Siri are now commonly used through mobile and smart speaker devices. These intelligent personal assistants (IPAs) use speech as the primary means of interaction, lacking embodiment compared to other speech systems such as robots or embodied conversational agents (ECAs). HCI research on these speech interface based intelligent personal assistants (IPAs) has looked to develop understanding of users’ expectations, perceptions and experiences. This work suggests that users tend to use human dialogue as an initial interaction metaphor, with expectations of system competence and function being set by the human-likeness of their design \cite{cowan_what_2017,cowan2017they,luger_like_2016}. Emulating aspects of humanness in speech interfaces is  common \cite{lee_social-psychological_2005,cassell2001embodied,clark2018social,cowan_voice_2015}, although the wisdom of this design choice has been questioned \cite{clark2018social, clark2016multimodal,moore_progress_2016, cowan_what_2017} as it may create unrealistic expectations of system capability \cite{luger_like_2016}. That said, currently it is not clear how people conceptualise the humanness of speech-based IPAs and how exactly users perceive that humans and speech interfaces vary when considering humanness as the metaphor for interaction. This knowledge is critical to informing the growing debate around humanness. By specifying the dimensions that users find important in this concept we can identify what type of design decisions may influence user expectations. Comparing humans and speech-based IPAs can also lead to a deeper understanding of the gulf between expectation and reality \cite{luger_like_2016, leahu_how_2013} when humanness is used as a heuristic for interaction.

The work presented contributes a user-centered insight into the dimensions that are important to users when considering humanness of speech-based IPAs and identifies how these vary from human dialogue. We do this by conducting a study using the Repertory Grid Technique (RGT) where, following interactions with three dialogue partners (a human, Amazon Alexa and Apple Siri), participants generated constructs that best described those interactions. Through content analysis of participant-generated constructs we identify eight salient dimensions along which humanness is most commonly conceptualised. These focus on \textit{partner knowledge set}, the qualities associated with \textit{interpersonal connections}, the \textit{linguistic content} of the partner’s utterances, a partner's \emph{performance and capabilities}, their \textit{identity and role} in the dialogue, the \textit{vocal qualities} they portray and \textit{behavioral affordances} toward a partner. We also outline the key differences between perceptions of humans and speech-based IPAs on these dimensions, with interview data showing users are highly aware of the limitations of IPAs in comparison and that they form a different category of communicative interaction along these dimensions. Our findings are the first to outline the multi-faceted nature of user perceptions of humanness, identifying the key dimensions that need to be considered when designing and researching humanness in speech interface interaction.

\section{Related Work}
\subsection{Humanness as an interaction metaphor}
Previous work on humanness has generally focused on embodied agents and robots, discussing concepts such as the uncanny valley \cite{mori19702012}. Studies have also explored categorizations of human-nature and human-uniqueness \cite{haslam2005more, haslam2006dehumanization} and how imbuing systems with personality traits increases perceived human likeness \cite{zlotowski2015persistence,ghafurian2019role}. Work also strongly emphasises how embodied factors mediate perceptions of humanness, showing influence from kinematics \cite{thompson2011perception}, agent appearance and embodiment \cite{breazeal2003emotion, schwind2018avoiding, schwind2018avoiding, ho2017measuring}, facial expressions and gesture \cite{bruce2002role, kuno2007museum}. Current, widely used IPA devices are not embodied beyond the device casing and lights that subtly change colour (i.e., they have no appendages, head or face, nor movement like a robot or avatar), so perceptions of humanness may not be driven by similar considerations. Yet as an overarching design principle for non-embodied speech-based IPAs, humanness is a fundamental concept in understanding interactions with current speech interfaces. In particular, speech system development focuses heavily on emulating human aspects of speech through expressiveness and human-like synthesis \cite{akuzawa_expressive_2018} or human-based conversational rules and structures \cite{edlund_pause_2009, gilmartin_social_2017}. This humanness seems to give users expectations as to the type of capabilities that a system may have as a dialogue partner \cite{luger_like_2016, leahu_how_2013}. Indeed, incongruence between expectations derived from humanness cues and the realities of interacting with speech-based systems is detrimental to speech interface user experience \cite{cowan_what_2017,luger_like_2016,moore_progress_2016,purington_alexa_2017}. Qualitative work investigating the user experience of speech-based IPAs for both power users \cite{luger_like_2016} and infrequent users \cite{cowan_what_2017}, found that they seem to initially use human dialogue as a metaphor for interaction. The humanness of current speech-based IPA design also leads to inaccurate expectations of system capabilities, which are adjusted after experiencing interaction \cite{luger_like_2016,leahu_how_2013}. This impact of humanness is echoed by \citet{moore_progress_2016} who highlights that human-like voices in particular may create overestimations of a system’s capabilities, resulting in potential communication breakdown and unsuccessful engagement with systems. Although the humanness communicated by the design of speech interfaces may act as an anchor for perceptions, people do seem to make a fundamental distinction between the capabilities of automated and human dialogue partners.  They see machines as more basic or functional in their conversational capabilities \cite{branigan_role_2011,clark_what_2019}, acting as “at risk” listeners in conversations \cite{oviatt_linguistic_1998}.

Understanding user’s perception towards speech interfaces may be important to unlocking reasons for people’s behaviours in interaction. Poor perceptions of system capability are assumed to cause us to adapt our language choices in interaction \cite{amalberti_user_1993,brennan1996conceptual,le_bigot_effects_2007,meddeb_what?_2010}. They drive us to use fewer fillers and coherence markers \cite{amalberti_user_1993}, more basic lexical choices, and shorter utterances compared to when we communicate with a human \cite{kennedy_dialogue_1988}, although the ubiquity of this role is debated \cite{cowan_voice_2015,cowan2015does}. 

\subsection{Understanding perceptions using repertory grids}
Understanding what factors influence these perceptions could be critical to understanding fundamental mechanisms in speech interface interaction. The RGT is a research tool designed to facilitate the discovery of what may drive users perceptions towards particular objects and situations. Developed as a tool to support personal construct theory in psychology, the technique allows participants to generate labels (termed \textit{personal constructs}) to describe, conceptualise and compare particular objects of study (termed \textit{elements}) \cite{kelly_psychology_1991}. Participants are exposed to three elements at a time, through a paradigm known as \textit{triading}, where constructs are generated to differentiate two similar elements from a distinct third element. RGT therefore allows researchers an insight into an individual's reasoning and conceptualising process as they work toward understanding their experiences \cite{fransella_manual_2004}. Historically, it has been used in educational psychology \cite{shaw_focus_1978} and information design \cite{hogan_blending_2013}. HCI-related studies have also utilized the RGT to examine perceptions of website usability \cite{tung_attributes_2009}, perceptions of strategic information systems \cite{cho2010exploring} and perceptions of mobile technologies \cite{fallman2010capturing}.

\begin{table*}[]
\caption{Question types with examples}
\label{table1}
\begin{tabular}{@{}cc@{}}
\toprule
\textbf{Question/request type} & \textbf{Question/request format} \\ \midrule
Conversational & \textit{\begin{tabular}[c]{@{}c@{}}How are you today?\\ Where are you from?\\ Tell me a joke\end{tabular}} \\ \midrule
Information retrieval & \textit{\begin{tabular}[c]{@{}c@{}}Who is {[}insert famous person's name{]}?\\ What is the square root of {[}insert three digit number{]}?\\ How do I get to the City Centre from here?\end{tabular}} \\ \midrule
Subjective/opinion-based & \textit{\begin{tabular}[c]{@{}c@{}}Do you like {[}insert favorite genre of music{]}?\\ Can you recommend a place to eat {[}insert favorite food when eating out{]}?\\ What do you think of {[}insert famous person's name - same as before{]}?\end{tabular}} \\ \bottomrule
\end{tabular}
\end{table*}

\section{Research Aims}
Humanness plays a critical role in the development of perceptions towards IPAs. Yet concepts of humanness are predominantly discussed in embodied interactions, as a unidimensional concept, with little known about how users conceptualise humanness in non-embodied speech interfaces. Insight into user-generated concepts of humanness is crucial to add nuance to the debate around its role in user experience. Mapping the dimensions that users find important when conceptualising the humanness of speech-based IPAs through a bottom-up construct elicitation paradigm will allow us to identify the design dimensions that may need to be prioritised to ensure accurate user perceptions are generated. Through using RGT, this study aims to: (1) map the dimensionality of the concept of humanness in interactions with speech-based IPAs; and (2) identify the nuances between human and IPA interaction within these dimensions. 

\section{Method}
Using random sampling, twenty-four participants from a European university were recruited via an email circulated across the university. Recruitment posters were also displayed around campus. Each participant was given a €10 honorarium for taking part. Three participants were omitted from the data due to them having difficulty in completing the grids unassisted. Of the remaining twenty one participants (F=9, M=11; Mean age = 23.1yrs; SD=5.49) all were native or near native English speakers. 23.8\% (N=5) of the sample reported using speech-based IPAs daily to a few times per week with 14.3\% of the sample (N=3) reporting only using speech-based IPAs a few times a month. 38.1\% of the sample reported using them rarely (N=8) with 23.8\% (N=5) having never used them before the study. Among those that had used speech-based IPAs, Siri (50\%) was most commonly used, followed by Google Assistant (31.3\%) and Amazon Alexa (18.8\%). 

\subsection{Procedure}
Upon arriving at the lab, all participants were fully briefed about the nature of the study and were asked to provide consent to take part. After completing demographics, participants were then given instructions as to the purpose of the study. Participants were informed that during the session they would have a brief interaction with three dialogue partners where they would ask each of them in turn a set of pre-determined questions (Familiarisation Phase). They were then told that after their interactions they would take part in a semi-structured interview where they would identify a series of constructs relevant to these interactions (Construct Elicitation Phase). After this, participants were asked to rate each interaction in relation to these constructs on a sliding scale (Rating Phase). The phases are described in detail below. 

\subsubsection{Familiarisation Phase}
Each participant completed a familiarisation session whereby they interacted with three elements. The elements selected included Siri (accessed on an iPhone SE smartphone), Alexa (accessed through an Echo Dot device), and a human (a member of the research team). The speech-based IPA elements were chosen because of their popularity and how they varied in their use of multimodality in interaction. Alexa, via the Echo Dot, predominantly uses speech, with minimal visual output through using a single coloured ring on the top of the device. Siri primarily uses speech as a primary interaction modality but also uses significant visual feedback through the smartphone’s touchscreen. The human was included to act as a comparator and ensure that constructs would relate to humanness. For ecological validity, human and speech-based IPA responses were unconstrained. Familiarization was completed with all three elements prior to the construct elicitation phase. The order in which elements were addressed was randomised between participants to control for recency and order effects. Only minimal introductions between the human and participants were made by the lead researcher to confine the interaction as much as possible within the predefined questions.

To facilitate interaction with each element, participants were tasked with asking each dialogue partner nine pre-set questions. These were devised by the authors to emphasise differences in the way humans and speech-based IPAs can communicate, so as to ensure a wide set of comparisons could be made. The questions used are shown in Table \ref{table1}. The order in which elements were addressed was randomised between participants. 

\subsubsection{Construct Elicitation Phase}
Next, through a semi-structured interview, participants were asked to compile a list of \textit{implicit constructs} that they felt best described the key similarities and differences between their interactions. These are critical constructs that inform the participant’s own understanding of their experiences \cite{kelly_psychology_1991,fransella_manual_2004}. Participants were asked to focus on the communicative abilities and qualities of each element. They were also asked to provide context and reasoning around why they were choosing certain words and how it related to their interactions. After an exhaustive list had been compiled (averaging 11-13 constructs), participants were asked to devise a dichotomous \textit{emergent construct} for each \textit{implicit construct} elicited. Participants were informed that this did not have to be a direct antonym - although it could be if appropriate - but should be based on the context in which they meant the original word. From this a grid of implicit-emergent construct pairs were gathered for each participant.

\subsubsection{Rating Phase}
In this final stage, participants were asked to rate where they felt each element fell between each of the implicit-emergent poles (construct pairs) elicited. To do this, participants were presented with a grid containing all constructs produced during the construct elicitation phase. Implicit constructs were entered into the left hand column, emergent constructs into the right hand column, and both columns were connected with a straight line, 134mm long. Participants were asked to place a mark on the line where they felt each element sat on a spectrum between the construct poles. This data was used to help interpret whether the implicit or emergent construct was more closely associated with human or IPA based experiences. They were also used to provide additional context and help interpret construct meanings, further supporting the categorisation process. 

All sessions were recorded and transcribed. Upon completing the session, users were debriefed as to the aims of the study and were thanked for taking part. 

\subsection{Analysis}
A bootstrapped content analysis of the construct pairs \cite{jankowicz_easy_2004} was conducted to identify consistent themes among the constructs generated. This categorisation helps identify overarching meanings or characteristics across the constructs \cite{jankowicz_easy_2004}. Following guidelines by \cite{jankowicz_easy_2004} the analysis was conducted over three phases. 

\subsubsection{Phase 1} Three of the authors with experience in qualitative and/or speech interface research independently assigned initial thematic tags to each implicit-emergent construct pair. This was based on semantic assessment of construct labels, using participant explanations from the interview data and ratings to ensure the participants’ intended meaning was captured. From this phase, an initial understanding and conceptualisation of common themes present in the constructs was produced.

\subsubsection{Phase 2} The themes from Phase 1 were then discussed by two of the annotators from the initial coding phase and two speech interface researchers who had not been involved in the initial coding, using the \textit{constant comparison method} \cite{dick_social_2001}. In this phase each construct pair was compared with each other to decide if they were similar. Further defining and refining of categories occurred as the process was undertaken to produce a final set of thematic tags. 

\subsubsection{Phase 3} As a final phase, two annotators (one of whom was not involved in Phase 2) used the themes generated in Phase 2 to independently annotate the data. Both of these annotators then discussed areas of disagreement and revised annotations independently. This process is important to reduce idiosyncrasies and improve the reliability of category identification \cite{dick_social_2001,jankowicz_easy_2004}. Agreement between raters was high [Cohen’s kappa = 0.76, p <.001]. A final discussion took place to resolve remaining disagreements before final annotations were performed by the lead author.  

\section{Results}
Participants across the study produced a total of 266 implicit-emergent construct pairs. The analysis resulted in eight key dimensions\footnote{A miscellaneous category was created when no agreement on category and construct explanation could be traced back to interview data} that categorise participants’ perceptions when comparing dialogue with a human to dialogue with speech-based IPA partners. A summary of the themes, selected construct pairs and total number of constructs linked to each theme are shown in Table \ref{Table 2:}. Some construct pair examples in the table were reversed so that constructs most closely related to the human partner appear on the left for ease of interpretation (e.g., Human/IPA). A full list of the constructs elicited in each category is included in supplementary material. Below we use data from interview transcripts to explore the definition, and differences between human and speech-based IPA dialogue experiences respectively.

\begin{table*}[]
\caption{Summary of construct analysis}
\label{Table 2:}
\begin{tabular}{clcc}
\hline
Theme & \multicolumn{1}{c}{\begin{tabular}[c]{@{}c@{}}Construct pair examples\\ (Ordered Human/IPA)\end{tabular}} & \begin{tabular}[c]{@{}c@{}}Number of construct pairs\\ (\% of all constructs generated)\end{tabular} & \begin{tabular}[c]{@{}c@{}}\% of agreement \\ between raters\end{tabular} \\ \hline
Partner Knowledge Set & \begin{tabular}[c]{@{}l@{}}Opinionated/Non-judgmental; \\ Biased/Neutral; Free-Bookish; \\ Expansive/Limited; \\ Ad-hoc/Spontaneous/Pre-programmed; \\ Colloquial/Universal knowledge; \\ Abstract/Specific knowledge; \\ Lateral/Inflexible thinking\end{tabular} & \begin{tabular}[c]{@{}c@{}}N=72\\ (27.1\%)\end{tabular} & 95.8\% \\ \hline
Interpersonal Connection & \begin{tabular}[c]{@{}l@{}}Personal relatability/Manufactured; \\ Genuineness/Ungenuine; \\ Real/Fake; Canny/Uncanny; \\ Emotional/Cold; Personal/Robotic; \\ Connection/Disconnected-disinterested; \\ Engaged/Remote; Humour/Humourless\end{tabular} & \begin{tabular}[c]{@{}c@{}}N=48\\ (18.1\%)\end{tabular} & 83.3\% \\ \hline
Linguistic Content & \begin{tabular}[c]{@{}l@{}}Short answers/Long answers; \\ Safe/Edgy; Expansive/To-the-point;\\ Convenience/Inconvenience; \\ Elaborate/Pointed; Polite/Blunt or rude; \\ Slang/Phrasing; Colloquial/Formal; \\ Vague/Detailed\end{tabular} & \begin{tabular}[c]{@{}c@{}}N=41\\ (15.4\%)\end{tabular} & 51.2\% \\ \hline
\begin{tabular}[c]{@{}c@{}}Partner Performance \\ and Capabilities\end{tabular} & \begin{tabular}[c]{@{}l@{}}Channels of communication Multiple/Single; \\ Speed of response: Slow/Fast; \\ Hesitating/Confident-decisive; \\ Recognition-Understandability (High/Low)\end{tabular} & \begin{tabular}[c]{@{}c@{}}N=21\\ (10.2\%)\end{tabular} & 63.0\% \\ \hline
Conversational Interactivity & \begin{tabular}[c]{@{}l@{}}Two-way/One-way; Conversive/Monologue; \\ Leading/Uninteractive; \\ Conjunctive/Disconnected; \\ Continuation/Stop \& Start conversation; \\ Conversational speech (fluid)/Stilted speech; \\ Keeping track/Isolated responses\end{tabular} & \begin{tabular}[c]{@{}c@{}}N=24\\ (9\%)\end{tabular} & 87.5\% \\ \hline
\begin{tabular}[c]{@{}c@{}}Partner Identity \\ \& Role\end{tabular} & \begin{tabular}[c]{@{}l@{}}Humanness/Machineness; \\ Real/Organic-Artificial; \\ Personalised/Commercialised; \\ No agenda/Agenda; \\ Transparency (purpose) (Low\textbackslash{}High); \\ To help/To serve; Level standing/Slave\end{tabular} & \begin{tabular}[c]{@{}c@{}}N=21\\ (7.9\%)\end{tabular} & 100\% \\ \hline
Vocal Qualities & \begin{tabular}[c]{@{}l@{}}Pronunciation/Incomprehensible; \\ Consistent/Choppy; \\ Pleasant/Unpleasant; \\ Audibility-Clarity of speech (High/Low); \\ Enthusiasm/Monotonous; \\ Speech modulation/No modulation; \\ Cheery \& Emotional/Dull \& Emotionless\end{tabular} & \begin{tabular}[c]{@{}c@{}}N=17\\ (6.4\%)\end{tabular} & 64.7\% \\ \hline
Behavioural Affordances & \begin{tabular}[c]{@{}l@{}}Natural/Conscious; \\ Familiarity/Unfamiliarity; Patience/Impatience; \\ Easy communication/Limited recognition \\ (make allowances for)\end{tabular} & \begin{tabular}[c]{@{}c@{}}N=13\\ (4.9\%)\end{tabular} & 84.6\%. \\ \hline
Miscellaneous & \begin{tabular}[c]{@{}l@{}}Fair/Unreasonable; \\ Calculated/General; \\ Personal/Impersonal\end{tabular} & \begin{tabular}[c]{@{}c@{}}N=3\\ (1.1\%)\end{tabular} & 100\% \\ \hline
\end{tabular}
\end{table*}

\subsection{Partner Knowledge Set}
72 of the constructs participants generated are focused on the perceived types of knowledge and information that each of the partners can supply during dialogue. Participants said that, unlike humans, speech-based IPAs’ knowledge tends not to extend to subjective opinions or judgements. The IPAs’ knowledge base was also perceived as more factual, whereas people were seen as relying on more experience based knowledge: 

\begin{center}
\emph{"...they provide very fact and evidenced based knowledge where people will give experience based knowledge" [P22]}
\end{center}

On occasions when a speech-based IPA did try to convey a subjective preference, responses were varied:  

\begin{center}
\emph{"Alexa’s opinion on Rock music, it said it liked Queen and I think Bohemian Rhapsody...felt very human-like" [P15]}
\end{center} 

Others felt these interjections made it feel like the IPAs were trying too hard to be human-like, and this served to highlight the role of the team who programmed the IPA to create an illusion of subjective knowledge on a particular theme: 

\begin{center}
\emph{"When Alexa made an X-Files reference I immediately thought of whatever Silicon Valley guy on his lunch break on a Tuesday was like 'Oh, I'm gonna make an X-Files reference'...it was almost trying too hard, it took me out of it"[P1]}
\end{center}

Essentially, participants saw IPAs' knowledge as objective \textit{“purely informative types of things”} [P3] that \textit{“just pull up a Wikipedia page...as quickly as possible”} [P1]. Conversely, humans were seen as having an ability to draw on richer forms of information meaning they could \textit{“talk for like ten minutes”} if asked \textit{“what do you think of David Bowie?”} or on a topic like \textit{“not even whether he was a good person but what is a good person”}[P1]. This leads to them being perceived as able to give \textit{“complex”} [P11,P22], and \textit{“personalised”} [P20,P21] answers to potentially complicated queries. Humans are seen as more able to contextualize, make use of socially relevant and colloquial knowledge, \textit{“make assumptions”} [P19] and draw supplementary information from the conversation as it unfolds: 

\begin{center}
\emph{"A big part of it is actually talking to someone who knows the greater context...to know the greater context of everything that’s going on around you, I guess that’s something the virtual assistants just can’t do right now" [P10]}
\end{center}

Humans were regarded as \emph{“spontaneous”} [P9], \emph{“nuanced”} [P14], and \emph{“interpretive”} [P22]. Yet, although speech-based IPAs were seen as having access to a greater \emph{“depth of information”} [P5], their \emph{“very bookish”} nature meant they were perceived as more \emph{“constrained”} [P5] and  \emph{“literal”} [P9,P22]: 

\begin{center}
\emph{“When I asked what [the human partner] thought about Chuck Norris she was saying like, ‘Oh, he’s very conservative’. The political aspect came into it...but say if Chuck Norris likes Pizza, well they [the IPAs] wouldn’t be able to say then maybe he likes cheese as well, maybe he likes tomatoes as well. They just know he likes Pizza” [P12].}
\end{center}


\subsection{Interpersonal Connection}
48 constructs generated emphasized perceived qualities associated with the development of interpersonal relationships between dialogue partners. A number of participants noted that, compared to the interaction with a human, interaction with both Siri and Alexa felt \textit{“fake”} [P10] or \textit{“completely false"} [P14]. This was sometimes amplified by \textit{“attempts to be funny (and) relatable”} [P14], that were interpreted as dry, boring and unfunny and generally seen as an attempt to simulate an artificial social connection. The issue of authenticity of the interaction also contributed to a sense of strangeness or eeriness. One participant remarked that they wanted these systems to have more humanlike qualities, but the eeriness of the current interpersonal interaction was also noted: 

\begin{center}
\emph{"One of the biggest things to me is like ‘is this thing humanlike?’ And then on the other side I'd say like uncanny...there were times when I felt I was sitting deep in the uncanny valley when I’m talking to these things"  [P10]}  
\end{center}

Some noted a lack of inherent emotion on the part of the IPAs when giving information, whereas it felt that humans generally express inherent emotion about a subject by default when conversing. Again the authentic nature of the interactional qualities were brought into question, where people felt that, although IPAs projected emotion at points, that \textit{“they were only pretending”} [P22]. 

Empathy and engagement were also prominently mentioned.  IPAs were described as largely devoid of the ability to show warmth compared to a human partner. The IPAs felt \textit{“quite disengaged compared to the human”} [P24], especially when responding to \textit{“personal questions”} like \textit{“where are you from? And how are you today?”} [P24]. Alexa’s tendency to \textit{“churn out answers”} [P24], meant it felt less \textit{“relatable”} and less \textit{“human-like”}. Likewise the IPAs’ \textit{“lack of feeling and lack of humor”} [P5] made them seem like \textit{“disinterested”} [P16] dialogue partners. Although some participants did characterise all the partners they interacted with as friendly, there were clear deficits in how relatable and interpersonal the IPAs were compared to the human partner.

\subsection{Linguistic content}
41 constructs generated focused on the structure and tone of linguistic content generated by each partner. IPAs' linguistic output was generally seen as formal, \textit{“always using full sentences”} [P4], along with being \textit{“polite”} [P20], \textit{“diplomatic”} [P24] and \textit{“safe”} [P11]. Between the IPAs, Alexa’s use of humor did make it seem less formal than Siri for some participants. However, human-like social content such as small talk was often seen as excessive, inappropriate, \textit{“obscure”} and \textit{“a bit weird”} [P19]. Indeed there was a strong mistrust when Alexa’s output involved opinion rather than formal fact based content.  For instance, one participant stated they felt \textit{“manipulated”} [P14] when Alexa expressed a favorite band and song. 

Participants also felt that human partners used a more narrative rather than descriptive style to convey responses. Although at points, IPAs were perceived as direct and to-the-point in their responses, Alexa was commonly criticized for providing too much information to queries compared to human partners and Siri. 

\subsection{Partner Performance \& Capabilities}
27 of the constructs were categorised as describing functional aspects of the way dialogue partners performed. Many of these emphasised the multimodal nature of communication and the functional limitations of IPAs. The most significant observations made by participants related to the importance of feedback through non-verbal communication in generating a sense of humanness. Specifically, they felt that communication through speech-only led to ambiguity as to whether their partner had recognised and understood what they had said. Siri’s use of visual feedback (e.g. presenting automatic speech recognition (ASR) output on screen) was particularly useful in reducing this ambiguity, and was compared to feedback given from human partners through cues such as facial expressions or hand gestures. Some felt Siri outperformed the human and Alexa in giving confirmation of understanding: 

\begin{center}
\emph{“Well yeah, everything was oral for Alexa, umm, and human, whereas Siri provided visual context as well...it's like a multifaceted approach" [P13]}
\end{center}

Others drew similarities between the efficiency of multimodal interactions with Siri and the use of  facial and hand gestures: 
\begin{center}
\emph{“She (Human) could give me a physical representation as an answer...something else that I thought was interesting was because you have like the phone, like when it gives you information it doesn't tell you, it just pulls it up...So like people can use various methods but also Siri uses pictures and stuff, whereas Alexa can't do that." [P22]}
\end{center}

Despite some issues with ambiguity, recognition and understanding were seen as strong across elements. Yet there were clear differences in how the precision and accuracy of information being delivered by each partner was perceived: 

\begin{center}
\emph{"If I were to take that (Human's) advice I wouldn't have known how to get to there because I wouldn't have known which bus to get, things like that" [P14]}
\end{center}

\begin{center}
\emph{“...the computers because they're automated don't make errors so there's a difference probably in the quality of information” [P24]}
\end{center}

The speed of response was also a clear point of divergence between the elements. Many participants noted that the IPAs were very quick to respond, whereas the human was sometimes hesitant: 

\begin{center}
\emph{“...there's like a lead time when you speak to a human...like the, 'Umm' 'Ehh'; where they're thinking like its a pause for thought, whereas a computer doesn't have the pause for thought, so I suppose, speed of response quick" [P24]}
\end{center}

\subsection{Conversational Interactivity}
24 constructs related to how conversational participants felt interactions with each of the elements were. A number of the constructs related to perceptions of the direction and interactivity of communication in these contexts with IPAs containing, \textit{“very little conversation and more, like commands”} [P10]: 

\begin{center}
\emph{"The human yep, very good conversation, very interactive, the others...it definitely didn't feel like it was a conversation. Like I was asking questions and waiting for a response" [P9]}
\end{center}

Participants perceived that the questions used in the study, when used with the human partner, were the start of a conversation that could unfold dynamically, rather than merely a request that was completed after an answer:

\begin{center}
\emph{“With the first question 'how are you today?', the human, kind of responded with 'yeah I'm good, how are you?', whereas, both Siri and Alexa were just an answer, and end of conversation” [P3]}
\end{center}

\begin{center}
\emph{“When I asked do you like rock music and she (Human) said, that depends what you mean by rock music, like that's the beginning of a conversation, it's not just an answer” [P1]}
\end{center}

This perceived lack of progression with IPA dialogue was consistently mentioned, with participants noting that to get more from the IPA you would have to \textit{“....ask them another question”} [P20]: 

\begin{center}
\emph{“With the human...when they give an answer, you can go on again. But with them (IPAs) it's like, once they give an answer, that's it” [P18]}
\end{center}

Likewise the fluidity of the conversational interaction with IPAs was an issue:

\begin{center}
\emph{“Some of the things they (IPAs) say is like stilted. Like, for example when she was telling that joke she's kinda like pausing, but that's not natural” [P18]}
\end{center}

\subsection{Partner Identity or Role}
21 constructs focused on the partner’s identity and the role they play within the dialogue. On the whole these constructs highlighted a simple dichotomy of partner identity between human and machine. There were also reflections on the commercial nature of IPAs and how this influences functionality:

\begin{center}
\emph{“...I'm aware of the company, because I know these are products sold for money” [P1]}
\end{center}

\begin{center}
\emph{“Like Google Assistant, I probably linked it to Spotify but I'm pretty sure the first time I asked it to play music it was like you can open this on Google Music” [P6]}
\end{center}

The power dynamic between user and speech-based IPAs also formed part of the theme’s constructs: 

\begin{center}
\emph{"Me and human are at the same level….when I'm (IPAs) asking the questions I don't really feel I'm asking the questions, I'm dictating for an answer...it's a master-slave type thing” [P14]}
\end{center}

On the other hand the IPA’s purpose was seen as very \textit{“transparent”} [P6] compared to humans who may have more complex motives and social roles within interaction.

\subsection{Vocal Qualities}
17 constructs emphasised aspects of non-lexical voice quality. This largely revolved around clarity of a partner’s voice. Participants felt that \textit{“their [IPAs] diction or their actual pronunciation was very sharp”} [P4], and that all partner’s voices were \textit{“...easy to understand. They're all comprehensible”} [P15]. This highlights improvements in speech synthesis and the development of clear and understandable voices. 

Clear differences between partners were seen when discussing voice expressiveness. Humans we were seen to \textit{“express emotions when we talk”} but \textit{“when they (IPAs) say something, they say it really coldly. Whereas...from a human it would have some ups and downs, or like some voice modulations that Alexa and Siri doesn't have”} [P19]. 

Some did note the attempt to include expressiveness in the voices used by IPAs, feeling that they were \textit{“...all relatively cheery. They didn't sound, in any way, dull...(but) the human was much more emotionally cheery I suppose”} [P16].

\subsection{Behavioural Affordances}
13 constructs focused on affordances people felt they needed to make to account for limitations of their dialogue partners. The vast majority of these types of observations related to participants feeling they had to adapt the way they spoke when interacting with speech-based IPAs by trying to \textit{“.. pronounce my words more accurately”} due to consciously considering \textit{“whether they understood”} [P21].  Participants also felt that they could be \textit{“more informal, or maybe more spontaneous”} [P9] when interacting with a human, and that queries made to speech-based IPAs were less natural and had to be structured in a particular way: 

\begin{center}
\emph{“I know how to ask a question to get the information that I want that wouldn't necessarily be the way you would converse with a human” [P10]}
\end{center}

This was a particular issue for some who felt interactions with IPAs had rules that needed to be learned: 

\begin{center}
\emph{“I wasn't familiar with like, how, to interact with them, I didn't know the rules” [P3]}
\end{center}

Finally, some participants suggested they would be more forgiving and patient in accommodating communicative limitations of humans compared to IPAs.

\section{Discussion}
Research \cite{cowan2017they,cowan_voice_2015,luger_like_2016,porcheron_voice_2018,dubiel_survey_2018,moore_progress_2016} has consistently identified that humanness is fundamental to speech interface design, supporting the metaphor of human dialogue for interaction. It is currently unclear what dimensions are important in users’ conceptualisation of humanness and how people see speech interface functionality through this lens. Through using the Repertory Grid Technique, we discover eight key dimensions which are the focus of humanness: 1) a partner’s level and type of knowledge or \textit{partner knowledge set}; 2) the potential for \textit{interpersonal connections}; 3) the \textit{linguistic content} they use in their dialogue turns; 4) the presence or absence of \textit{conversational interactivity}; 5) functional attributes that relate to a \textit{partner’s performance and capabilities} to complement dialogue; 6) a coarse grained judgement of their \textit{identity and role} in the dialogue, 7) the expressiveness and clarity of their \textit{vocal qualities} in the speech they produce and 8) \textit{behavioural affordances} made to accommodate for perceived limitations of dialogue partners. These dimensions, based on categorisation of user-generated constructs, echo previous researcher led attempts to map fundamental differences between humans and machines \cite{haslam2006dehumanization} (e.g. ‘interpersonal connection’ is similar to ‘emotional responsiveness vs inertness’, and ‘partner knowledge set’ mirrors ‘cognitive openness vs rigidity’).

Although people may use humanness as an anchor \cite{luger_like_2016,cowan2017they} people make clear distinctions between humans and speech-based IPAs on these dimensions. Participants see IPAs as being more fact based than opinionated in the knowledge they have, with human partners being able to combine knowledge and ideas in novel ways to contribute to dialogue. Attempts made by speech-based IPAs to be socially relatable and to create a connection were regularly seen as fake, with IPAs seen as lacking in emotion or perceived interest in the dialogue. IPAs were also perceived as far more formal and verbose in the volume of content they deliver, with social or opinion based content being viewed with suspicion. Differences in the levels of feedback between humans and IPAs were also expressed, whereby speech-only devices were limited when compared to partners that use visual or other multimodal signals to convey understanding. Speech-based IPAs were judged as more able to return accurate and faster responses compared to human partners. Whilst human dialogue was two way and more expansive in nature, IPAs by comparison seemed limited to question-answer type interactions. Above the clear human/machine dichotomy participants also perceived that IPAs were built to serve the user, with humans having more complex social roles and purposes when conversing. Although voice clarity did not vary between partners, there were clearly perceived differences in expressiveness and paralinguistic features, whereby the speech-based IPAs seemed cold and less emotive than a human. Participants also noted a perceived lack of flexibility and a need to understand the rules of interaction when communicating with speech-based IPAs, with humans seen as more adaptable to how the user may want to converse.

\subsection{The multi-dimensional nature of humanness}
Our major contribution is in identifying the key dimensions that users use to conceptualise humanness and how these are mapped to speech-based IPA interactions. This adds much needed richness to the view of humanness as an influencer of the speech-based IPA user experience \cite{cowan_what_2017,luger_like_2016}, by breaking down the concepts, identifying important dimensions, and identifying the similarities and differences in how these are applied to humans and speech-based IPAs. From our findings it is clear that humanness needs to be considered not as unidimensional, but as a multi-dimensional concept in speech interface design. The primary purpose of the analysis was not to generate design recommendations, rather it sets to identify the dimensions that influence perceptions of humanness and how they differ in human and IPA interaction. That said, designers could use these dimensions to inform decisions and areas of focus. For instance, our themes could aid in understanding how users frame their experience when compared to human interaction, thereby identifying pros and cons of using humanness as a design metaphor. Striving for humanness is a common guiding principle in speech interface design \cite{aylett2019siri} and a heuristic users frame their experiences through \cite{cowan_what_2017,leahu_how_2013, luger_like_2016}. A nuanced understanding of how users comprehend their interactions through this heuristic is valuable to inform design decisions. With regards to research, a number of themes do indeed echo topics within speech interface research. For instance, vocal quality clearly maps to work on speech synthesis, where developing more human-like, expressive \cite{akuzawa_expressive_2018}, emotive \cite{meddeb_what?_2010} and personality-filled \cite{wester_artificial_2015} voices is currently underway. User research has also focused on exploring the role of humanness in partner knowledge assumptions \cite{cowan2017they}, vocal quality \cite{cowan_voice_2015,lee_social-psychological_2005}, partner identity \cite{branigan_role_2011}, linguistic content \cite{clark_what_2019} and conversational interactivity \cite{clark_what_2019,purington_alexa_2017}. Our work not only identifies the important dimensions of humanness when considering user perceptions, but formalises intuitions by researchers on the types of design decisions that may affect these perceptions.  
\subsection{Issues with emulating aspects of humanness}
Our findings show that humans and speech-based IPAs are clearly perceived differently on the dimensions identified. IPA interaction is seen as predominantly unidirectional and as a master-servant relationship compared to human-based dialogue interaction. They are also perceived as more socially fake and limited in their interpersonal ability than human partners, as well as seeming more formal and specific than human interlocutors. 

This supports the notion that, although they may be perceived through the lens of human dialogue, speech interface dialogues are fundamentally distinct \cite{porcheron_voice_2018, clark_what_2019}. Currently, dialogue interactions with speech-based IPAs tend to be isolated question-answer pairs \cite{porcheron_voice_2018}, seemingly functional rather than interpersonal or social in nature \cite{clark_what_2019}. It may be argued that this is because of truly natural aspects of dialogue not being implemented appropriately or at all in the types of devices currently available. Especially in terms of interpersonal connections and conversational interaction, current projects are exploring how to include this in speech interface design \cite{spillane_introducing_2017,devillers2018multifaceted,fang2017sounding,papaioannou2017alana} meaning this may become more common and implemented more effectively. Our data suggests some fundamental barriers to accepting such an interaction. Small talk was sometimes flagged as inappropriate or undesirable for current speech-based IPA partners. More social features were  perceived as inauthentic, owing to the perceived pre-programmed nature of these systems, rather than because of the types of interactions speech-based IPAs were designed for. Although context and the purpose of the  interaction may play a role in the acceptability of social functionality, this type of functionality may be pushing the fundamental limits of spoken interactions between humans and machines \cite{clark2018social,moore_progress_2016}.

That is not to say that certain dimensions of humanness identified in our work should not be considered in future speech interface design. Concentrating on other dimensions of human-human communication may be more appropriate and even beneficial to user experience for speech interfaces interactions, including speech-based IPAs. In particular for IPAs, our participant highlighted making minor adjustments towards more human-like ways of delivering linguistic content,  like favouring brevity in message delivery and including multimodal feedback available in dialogue. This may improve the interaction experienced. Aspects such as developing expressive vocal qualities may also be useful in more accurately communicating emotive content through synthesis. 

\section{Limitations \& Future Work}
Because of the aim of the work, and the triading paradigm inherent to RGT, participants were asked to make direct comparisons between a human and two speech-based IPAs using the human as comparator. This meant that the human was highly salient in the comparisons being made between the three elements, potentially influencing which differences are prioritised. The purpose of this salience was to elicit dimensions and constructs particularly focusing on how humanness may be framing dimensions that people consider in speech-based IPA dialogue. This was so as to reflect findings in the literature that emphasise the important of humanness in design and the role this may have in mental model development \cite{luger_like_2016,cowan2017they}.
Future work could include more elements to limit the impact that each may play in guiding user constructs.

Our study involved participants reflecting on their interaction with the three elements. This interaction revolved around nine pre-determined questions. This technique was useful to expose participants to particular interaction issues and to ensure all participants could reflect on similar experiences when eliciting constructs. Yet it may have limited the types of dialogues participants reflected on, especially when interacting with the human dialogue partner. Future work may look to use more free-form interactions with all partners, which may not only replicate the categories for constructs found in this research, but add to their richness. 

It is also important to note that our study gives a snapshot of user perceptions on these dimensions, yet does not provide information as to how these may be fluid over an interaction, change with different experiences, or how these may change from initial perceptions before interaction. Mapping this dynamism has been noted as a consistent challenge in partner model and perspective taking research, in both human \cite{branigan_role_2011} and machine \cite{cowan_voice_2015} dialogue. Future approaches to this topic should focus on novel methods to gain a view on the dynamics of user perceptions. The themes identified may also act as a starting point to develop a measure to allow researchers and designers to quantify the impact of design choices on the dimensions of humanness. Validated metrics to quantify user perceptions are needed in the field \cite{clark2018state}. We plan to use this work as a starting point for developing such a metric.

\section{Conclusion}
Concepts of humanness are core to the design of speech interfaces like IPAs, yet the specific dimensions of humanness that people use to define these interactions are not fully understood. Our study has clearly outlined key themes related to how users view humanness in dialogue interaction and how this varies in speech-based IPA dialogue. It highlights that whilst humanness may be integral to speech interface design, significant thought needs to be placed into how humanness may be achieved and implemented with sensitivity to the specific dimensions identified.

\begin{acks}
This research was supported by an employment based PhD scholarship funded by the Irish Research Council and Voysis Ltd (R17830). 
\end{acks}

%
\balance
\bibliographystyle{ACM-Reference-Format}
\bibliography{RepGrid.bib}

\end{document}